\documentclass[journal]{IEEEtran}
\ifCLASSINFOpdf
\else
\fi
\usepackage{epsfig}
\usepackage{amsmath}
\usepackage{algorithm}
\usepackage{bm}
\usepackage{cite}
\usepackage{amsfonts,amssymb}
\usepackage{booktabs}
\usepackage{url}

\begin{document}
%
\title{FamDroid: Learning-Based Android Malware Family Classification Using Static Analysis}

%
%
%

\author{Wenhao~Fan,~
	    Liang~Zhao,~
	    Jiayang~Wang,~
	    Ye~Chen,~
	    Fan~Wu,~
	    Yuan'an~Liu
\thanks{The authors are with the School of Electronic Engineering and Beijing Key Laboratory of Work Safety Intelligent Monitoring, Beijing University of Posts and Telecommunications, Beijing, China. Liang Zhao and Jiayang Wang are the joint second authors since they contribute equally in our work. Email: whfan@bupt.edu.cn}
\thanks{Manuscript received XXXX XX, XXXX; revised XXXX XX, XXXX.}}

%
%

\markboth{XXXXX,~Vol.~XX, No.~X, XXXX~XXXX}%
{XXXXX \MakeLowercase{\textit{et al.}}: XXXXX}
%



\maketitle

\begin{abstract}
Android is currently the most extensively used smartphone platform in the world. Due to its popularity and open source nature, Android malware has been rapidly growing in recent years, and bringing great risks to users' privacy. The malware applications in a malware family may have common features and similar behaviors, which are beneficial for malware detection and inspection. Thus, classifying Android malware into their corresponding families is an important task in malware analysis.
At present, the main problem of existing research works on Android malware family classification lies in that the extracted features are inadequate to represent the common behavior characteristics of the malware in malicious families, and leveraging a single classifier or a static ensemble classifier is restricted to further improve the accuracy of classification.
In this paper, we propose FamDroid, a learning-based Android malware family classification scheme using static analysis technology. In FamDroid, the explicit features including permissions, hardware components, app components, intent filters are extracted from the apk files of a malware application. Besides, a hidden feature generated from the extracted APIs is used to represents the API call relationship in the application. 
Then, we design an adaptive weighted ensemble classifier, which considers the adaptability of the sample to each base classifier, to carry out accurate malware family classification. We conducted experiments on the Drebin dataset which contains 5560 Android malicious applications. The superiority of FamDroid is demonstrated through comparing it with 5 traditional machine learning models and 4 state-of-the-art reference schemes. FamDroid can correctly classify 98.92$\%$ of malware samples into their families and achieve 99.12$\%$ F1-Score.
\end{abstract}

\begin{IEEEkeywords}
Android security, malware classification, API call, ensemble learning, static analysis
\end{IEEEkeywords}

%
\IEEEpeerreviewmaketitle

\section{Introduction}
Nowadays, we can apply smart phones to nearly all aspects of our lives, such as information retrieval, online payment, positioning and navigation, multimedia entertainment, etc.
The global mobile OS market share report from International Data Corporation (IDC)  indicates that Android has become the most widely used mobile operating system \cite{idc_report}, which has a market share of 85$\%$ as of the third quarter of 2020. Meanwhile, Android malware grows explosively that seriously endangers the privacy security of users.
McAfee’s mobile threat report stated that as of the first quarter of 2019 \cite{macfee_report}, there has been 30 million mobile malware, and there is still a lot of malware emerging. In order to keep the Android market safe and protect every user's privacy, it is urgent to develop efficient detection tools that can curb the spread of Android malware.

Although many methods have been proposed to classify Android applications as malicious or benign, it is equally important to divide Android malware into the families they belong to. Since most of the new Android malware applications are variants of known malware families, we can analyze the unknown malware by predicting its family. The applications from the same Android malware family are inherently related. Therefore, Android applications from the same malicious family may share the similar code and perform similar malicious behaviors. Classifying Android malware applications into various known families not only can make an improvement on the effect of malware detection but can find the regular characteristics from the behaviors of malware that are helpful to analyze the unknown malware. 

There exist many successful schemes for Android malware family classification which achieved significant effects in different ways \cite{dendroid, droidminer, droidsift, ec2, AOM, MVIIDroid, fm, C-DFG, FalDroid, PST, HRL, A3CM}. However, most of these works fail to take into account the representative characteristics used to describe the common malicious behaviors in each family. In the feature extraction phase, the explicit features, which are widely used in existing works, such as \cite{ec2, MVIIDroid, fm, PST,HRL, A3CM}, can be extracted directly from the decompiled application files. But some malicious behaviors of a malware family are usually hidden in the implicit features which need to be obtained after a series of operations and transformations. These hidden features can reflect the characteristics of malware families. Leveraging these implicit features that represent more comprehensive and valuable behavioral characteristics can make an effective improvement on Android malware classification.
In the training phase, single classifier or an ensemble classifier is commonly used among the existing schemes, such as \cite{ droidsift, ec2, AOM, fm, C-DFG, PST, HRL, A3CM}. Different from the binary classification of malware detection, the multiple malware family classification is more complex. Generally, compared with ensemble classifiers, a single classifier has weaker generalization ability, and may cause local optimum while using a large number of samples. Ensemble learning combines multiple single classifiers, and it exploits the advantages of these classifiers to improve the performance of classification. But for a common ensemble classifier, the weights of all base classifiers are fixed. It is harder to keep a high accuracy while making the classification of unknown malware.

To this end, in this paper, we propose FamDroid, a learning-based Android malware family classification scheme using static analysis technology. In the feature extraction phase, the basic features including  permissions, hardware components, app components and intent filters are extracted from the AndroidManifest.xml and smali files. Meanwhile, the API call relationship, a type of implicit features which consists of the combination of multiple APIs, is extracted from the explicit APIs in malicious applications with a series of transformations and operations. In the feature selection phase, we combine different types of features and evaluate the importance of all features through the Random Forest algorithm \cite{random_forest}, then transform the feature vectors into low-dimensional feature vectors. In the phase of machine learning, we design an adaptive weighted ensemble classifier for Android malware family classification. In the training phase, an improved K-means algorithm algorithm is designed to cluster the training samples, and then form the base classifiers according to the clusters. In the testing phase, the weight of each testing sample to each basic classifier is adaptively assigned. In the experiments, we use the Drebin dataset \cite{drebin} to validate FamDroid. Compared with 5 traditional machine learning models and 4 state-of-the-art reference schemes, FamDroid achieve the best classification effects among them. 

The main contributions we made in this paper are as follows:

{$\bullet$} FamDroid considers the call relationship between different APIs in malicious applications and makes a good description of the correlation between the composite API calls and malicious behaviors. For malware, an API with high importance for classification indicates that there exists a certain behavior closely related to the calling of this API. 
In common, a complicated malicious behavior needs multiple key APIs to call each other. So we find a type of implicit features, the API call relationship, which can fully recapture malicious behaviors. To obtain the API call relationship, the relatively useless APIs are filtered out first and the rest of valuable APIs are used to generate API call graphs. We use a special matrix and assign the values of the matrix to represent the call relationship of the key APIs by analyzing the API call graphs we obtained.

{$\bullet$}  FamDroid considers the similarity between each testing sample and each base classifier and designs an adaptive weighted ensemble classifier. We improve the K-means algorithm to cluster the training samples scientifically by adding the weights to each training sample. In addition, we adaptively assign the weight between a base classifier and a testing sample by calculating the distance from a testing sample to the center of a base classifier. Different from the classical ensemble learning, each base classifier can achieve a properest weight while entering a new testing sample, which reduces the probability of misclassification.

{$\bullet$} The experiments are conducted on the Drebin dataset, which is the largest open Android malware data set with tag families. It contains 5,560 Android malware applications of different families to validate the feasibility of our scheme. To compare the classification effects under different feature subset, we select the most effective feature subset with a proper feature number. Compared FamDroid with 5 traditional machine learning models including Support Vector Machine(SVM), Logistic Regression(LR), Gradient Boost Decision Tree(GBDT), Naive Bayes(NB) and K-Nearest Neighbor(K-NN) and 4 state-of-the-art reference schemes, \cite{MVIIDroid, AOM, C-DFG, FalDroid}, the experiment results demonstrates that FamDroid has obvious superiority in classification, which can achive 98.92$\%$ accuracy, 99.23$\%$ macro precision, 99.17$\%$ macro recall and 99.12$\%$ macro F1-Score.

The rest of this paper is organized as follows: Section 2 shows the existing related work of Android malware family classification. Section 3 introduces the framework of FamDroid in detail. Experiments and comparisons are described in section 4. Finally, our works are concluded in section 5.

\section{Related Works}
Facing the threat posed by malicious Android applications, an increasing number of researchers all over the world have conducted research works on how to detect malicious software. In this section, we discuss the main existing works on Android malware family classification.

To achieve better performance on detecting malware, some research works that focussed on mining information from the code, considered a fine-grained family classification of Android malware. Dendroid \cite{dendroid} proposed a text mining method to analyze the code structure of Android malware and classified Android malware into families based on the similarity of the code structures. Droidminer \cite{droidminer} used static analysis to automatically mine malicious program logic from known Android malware by transforming it into a series of threat patterns, and then performed Android malicious application detection and Android malware family classification based on whether the application contained corresponding threat patterns.
Droidsift \cite{droidsift} was based on extracting features from the graphs and constructing dependence graphs to classify Android malware. The authors classified Android malware families based on the graphic similarity between known Android applications and unknown applications.

Among the existing schemes, some new and outstanding methods perform well on malware classification. They focussed on extracting different types of features and adopted different models for classification. The authors in \cite{PST} proposed a scheme that used common behavioral characteristics to classify malware. The scheme used the carefully chosen features commonly appeared in the same family to measure the similarity between malware families and adopted a community detection algorithm. A new Android hybrid representation learning approach is proposed in \cite{HRL} to cluster the weakly-labeled Android malware which can be partitioned into known and unknown families.
The authors in \cite{A3CM} created a new ground truth dataset that consists of 6,899 annotated Android malware samples from 72 families and used the multi-label classification model to annotate the malicious capabilities of the candidate malware samples. EC2 \cite{ec2} proposed an Android malicious application family classification algorithm based on a supervised classification algorithm and an unsupervised clustering algorithm. This scheme can effectively divide Android malware samples into malicious application families of different sizes. MVIIDroid \cite{MVIIDroid} used a Multiple View Information Integration Approach for Android malware detection and family identification. They extracted permissions, APIs, Dalvik opcodes, and native library opcodes as multiple type features and used the Multiple Kernel Learning algorithm as the classification model.  Android-oriented Metrics \cite{AOM} utilized a set of Android-oriented code metrics for static detection and used the Random Forest model as the classifier. In \cite{C-DFG}, the authors extracted features from control-flow graph and data-flow graph, then built the family classification models via deep learning. In \cite{fm}, the authors extracted features from source code and manifest files, including system API calls, hardware, permissions, and Android components, then used the factorization machine classifier as the final classifier for Android malware detection and Android malware family classification. A weighted-sensitive-API-call-based graph matching approach and a scheme called FalDroid are proposed in \cite{FalDroid} to handle the familial classification of large-scale Android malware. Malicious Applications from the same malware family have common behaviors which are conducted through multiple calls of APIs. However, most of the existing works only consider the explicit features that are difficult to represent common malicious behaviors in a malware family. Moreover, with the growing number of the new malware, normal classifiers have weak adaptability to them so that the performance of Android malware classification is constrained.

\section{System Model}
\subsection{Overview}

\begin{figure*}[t]
	\center
	\includegraphics[width=1\linewidth]{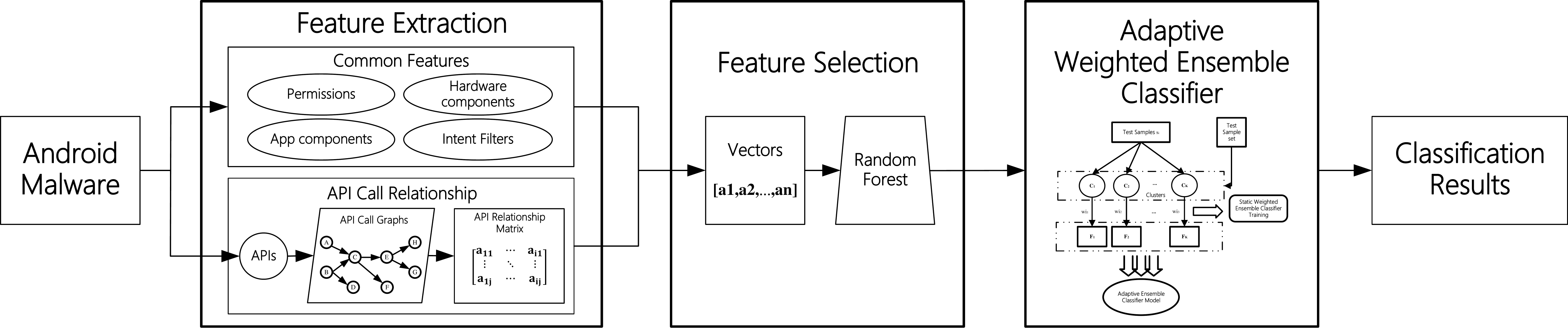}\\
	\caption{The framework of the proposed scheme}
	\label{Fig.1}
\end{figure*}

The overall framework of FamDroid is shown in Fig. \ref{Fig.1}. 
The classification scheme is divided into three steps: feature extraction, feature selection, and adaptive weighted ensemble learning.

(1) In the feature extraction phase, we collected Android malware samples from the Drebin dataset \cite{drebin} and decompiled these apk files into manifest files and smali files. We extracted 5 specific features from the decompiled files including 4 types of explicit features: permissions, hardware components, app components, intent filters and one type of implicit features: API call relationship. The explicit features can be obtained directly from the decompiled files. For the implicit features, we extract the API call relationship by analyzing the combination of different API calls. The relationship of API calls can be reflected by the links between valuable APIs. We traverse all opcode sequences from smali files and match the key APIs and their connections in the statements to generate the API call graphs. After the decomposition of the graphs, we can obtain the links that represent the relationship of API calls and express them via a relationship matrix.

(2) In the feature selection phase, we use the Random Forest model \cite{random_forest} to measure and rank the importance of each feature. The importance value is calculated by randomly adding the noise and then observing the change of the error prediction of the out-of-bag. According to the importance value calculated by the Random Forest, the irrelevant and redundant features with low importance value are removed from the feature set. Meanwhile, an effective feature subset with high importance value is selected. 

(3) In the ensemble learning phase, we design an adaptive weighted ensemble learning algorithm based on the clustering algorithm and the static weighted ensemble learning algorithm. In the training phase, the training samples are divided into several clusters by the improved K-means algorithm which considers the feature importance calculated by the Random Forest to generate the weight of each dimenssion. In the testing phase, through finding the distances between the testing samples and the base clusters, the weight of each testing sample to each basic classifier is adaptively assigned. To divide the dataset into the training set and the testing set, 5-fold-validation-method is adopted.

\subsection{Feature Extraction}

It is essential to obtain the effective features that can easily distinguish Android malware from different families. We convert the apk files with the labels of malware into manifest files and smali files by using the Apktools\cite{apktools} which is a famous decompiling tool for Android apk files. There are four types of explicit features and a type of implicit features. These different types of features can be extracted from manifest files and smali files as follows:
\subsubsection{Permissions}
The permission is an important part of the Android system security. At the beginning of the installation process, a list of all the permissions used in the application should be launched in the AndroidManifest.xml. The approval of permissions involves the invocation of various resources in the scheme. The invocation of some sensitive resources requires a specific permission declaration or a combination of permissions. Different families of Android malware usually have different malicious motives, so they usually apply for different permissions for specific malicious purposes. For example, a malware family, the MobileTx, will apply for the $RESTART\_PACKAGES$ permission to kill antivirus and the application monitoring systems.
\subsubsection{Hardware Components}
In the Android manifest file, the hardware that might be used by the application also need to be declared. The declaration of hardware is closely related to application functions. For example, some Android malware of the privacy disclosure family usually requests both the camera and network communication. It may lead to the leak of the users' private photos or videos.

\subsubsection{App Components}
Android components are inseparable from the implementation of program functions. The 4 major components of Android include: Activities, Services, Broadcast receivers and Content providers. Each component needs to be registered before being used. Triggering the malicious functions, running the service and reading the data are all related to the Android components. For example, the malware family, DroidKungFu, shares the name of some specific services\cite{drebin}. Therefore, collecting various components of an application is useful for identifying malware and classifying malicious families.
\subsubsection{Intent Filters}
The Intent Filters in malware can be used to request the components of applications to perform specific actions that endanger the privacy security of users. Each Intent Filter specifies the type of Intent it accepts according to the Action, Data and Category of the Intent. In malware, some implicit intents are related to the malicious behaviors. In order to indicate the implicit Intent that can be received by the application, one or more IntentFilters should be declared for the corresponding application component in the manifest file. Some malware families often use some specific intent filters, such as $SIG\_STR$, $Boot\_completed$, to perform malicious acts or illegally collect certain information.

\subsubsection{API Call Relationship}

The functions of applications are implemented through a large number of Android APIs. However, the APIs obtained directly from the smali files are not representative enough to reflect the comprehensive functionality of the applications related with malicious behaviors. A typical malicious function is normally implemented through calling a fixed combination of a few APIs. The hidden relationship of the key APIs can represent the further characteristic information inside applications. Therefore, we consider using the API call relationship which contains the key APIs and the links between them to distinguish the category of applications. The concrete process of getting the API call relationship is as follows:

During the process of preparation, the APIs relatively beneficial to our categorization are selected. Some of the APIs occur in almost all malware and some of the APIs seldom exist in malware. Therefore, the relationship between these API calls are not helpful for malware classification. Since these useless APIs require a large amount of computing resources and decrease the accuracy and efficiency of the classification model, it is necessary to remove them. We adopt the Random Forest which is derived from the construction of ID3 Decision Tree, to filter out these useless API features. Its decision-making is based on the information gain, which first obtains the value of the attribute set, and then uses the value to select the instance category.
The information gain is defined as the amount of information that the feature provides to the category label. To our experiments, it is an efficient feature selection method. The formula for the information gain is shown below:

\begin{equation}
	Ig(C,f) = H(C) - H(C|f)
\end{equation}
where $H(C)$ is the information entropy that represents the Android malware family category label $C$, and $H(C|f)$ is the conditional entropy that represents malware family category label with the given features. By calculating the information gain of every feature, we retain the features of higher importance and make them into a new vector.

After selecting the APIs, we design an algorithm to generate the Android API call graph and make smali files as the input. The algorithm is implemented through a series of steps. The opcode sequences are extracted from smali files first and then the invoke statements are matched. We traverse all invoke statements to record the call relationship between APIs and generate the API call graph. The detailed process of this algorithm is shown in $Algorithm 1$.

\begin{algorithm}
	\caption{Process of generating an API call graph from a decompiled application $S_{n}$}
	{
		\bf INPUT: smali files of $S_{n}$ }\\
	1. form our selected API list $A$\\
	2. randomly select an API as the initial node $n_{0}$\\ 
	3. extract the opcode sequences from smali files\\
	4. \textbf{for} each statement $s_{i}$ in the opcode sequences \textbf{do}\\
	5. ~~~~\textbf{if} $s_{i}$ is an API invoke statement \textbf{then} \\
	6. ~~~~~~~~retain  $s_{i}$\\
	7. ~~~~~\textbf{else}  \\
	8. ~~~~~~~~ delete $s_{i}$\\
	9. ~~~~\textbf{end if} \\
	10. \textbf{end for} \\
	11. \textbf{for} each $s_{i}$ retained  \textbf{do}\\
	12. ~~~~\textbf{if} $s_{i}$ starts with the node $n_{0}$ and $n_{0}$ calls another API which is in our selected API list $A$ \textbf{then} \\
	13. ~~~~~~~~create a new node $n_{j}$ and create the links between the two nodes\\
	14. ~~~~\textbf{end if} \\
	15. \textbf{end for} \\
	16. \textbf{return} $n_{0}$ \\
	17. traverse all nodes in the graph and execute the above steps\\
	{\bf OUTPUT: A directed API call graph }
\end{algorithm}	

\begin{figure}
	\center
	\includegraphics[width=5cm]{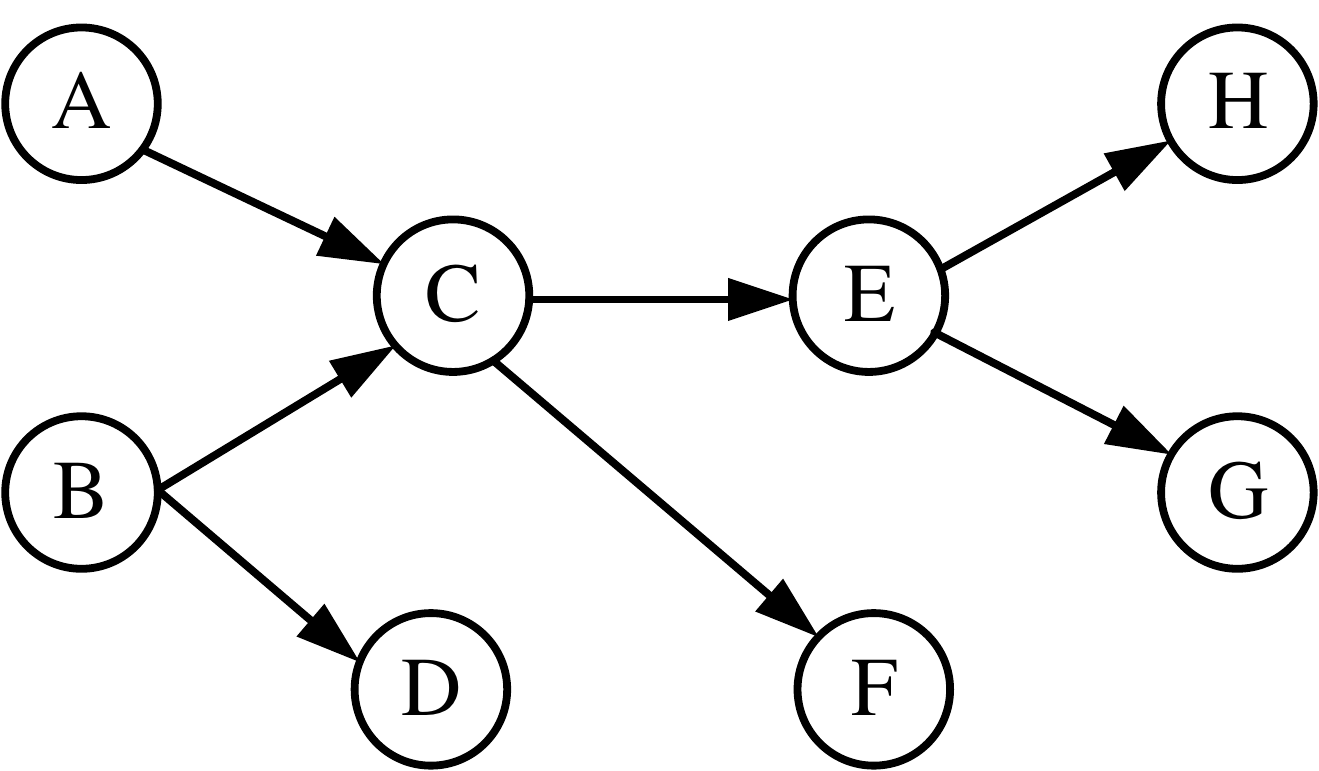}\\
	\caption{API Call Graph}
	\label{Fig.2}
\end{figure}

The API call graph shows the complete process of calling the key APIs of an application. In the graph, each node represents a key API which occurs in the opcode sequences and is requested by the malicious behaviors. A node in the graph usually has more than one branch, which means the different conditions for program execution. We use arrows to describe the links between different API calls and indicate the calling direction between an API and another API called by the prior API. As shown in Fig. \ref{Fig.2}, the nodes from $A$ to $H$ represent 8 different APIs.  We can observe how an API call another API and find the call relationship between these APIs from Fig. \ref{Fig.2}.  

To clearly reveal the API call relationship in an application, We extract the API call relationship from the graphs and adopt a specific structure to store this kind of features. The matrix is a two-dimensional structure which is suitable for storing the information about the connections between APIs. 
To get the matrix about API calls, we accomplish the traversal of the API call graph by traversing all nodes of the graph and making statistics of the links between all nodes until going out of the last node. We generate the API relationship feature matrix as follows:
\begin{equation}
	M=\begin{pmatrix}
		m_{11} & m_{12} &  \cdots & m_{1n}\\ 
		m_{21} & m_{22} &  \cdots & m_{2n}\\ 
		\vdots & \vdots &  a_{ij} & \vdots\\ 
		m_{n1} & m_{n2} &  \cdots & m_{nn}
	\end{pmatrix}
\end{equation}
Each element represents the call relationship between 2 specific APIs and the categories of APIs depend on the location of the element. The value of an element is 1 indicates that there exists a path between these 2 APIs and 0 indicates that these 2 APIs needn't call each other. From the perspective of the program code, the APIs that implement the same function are often called together. In the same way, there may be no functional association between the two APIs if there exits no linked path between them. Through this method, the information of the API call relationship of an application is stored in this matrix. 

Finally, We convert all features into a vector, the feature vector we make is composed of 0s and 1s, with 1 indicating that the malware has this feature and 0 indicating not. In this way, we transform combination of the feature sets into a $|S|$-dimension vector, where set $S$ is the set containing all features.

\subsection{Feature Selection}
The main purpose of this step is to select the most effective feature subset from the feature set. Through the above steps, we obtain a feature set that combine 4 explicit types of features and one implicit type of features. However, there exist a large amount of features in the set that are not related to the malicious functions and are not useful to categorize malware family. Some of extracted features are widely distributed in applications of all malware family, and some features appear individually and erratically. If these worthless features are not removed, the high-order feature vectors will require enormous computing resources and lead to the inefficiency of the classification. To achieve a high accuracy of classification and improve the efficiency of our scheme, it is necessary to conduct feature selection on the feature set to filter out redundant or irrelevant features and retain valuable features.

In this paper, we use the Random Forest\cite{random_forest} to finely select the features. The Random Forest is a popular and effective algorithm for the feature selection and multiple classification. It is based on the idea of model aggregation and is used for classification and regression problems. Unlike most other classifiers, the Random Forest can also effectively evaluate the importance of individual features and rank them in order of importance while conducting classification tasks. Despite the data used to generate the decision trees, there remains part of the data called  out-of-bag (OOB) to measure the accuracy of the classification. The Random Forest calculate the value of the importance of each feature by the following steps:

(a) For each decision tree, select the corresponding OOB to calculate the error of the classification and record it as $e_{1}$.

(b) Select the feature $f$ in all OOB, add noise interference by randomly changing the value of the feature vector, and then calculate the OOB error again and record it as $e_{2}$.

(c) For a forest with N trees, we mark $N_{t}$ as the the number of decision trees in the Random Forest. The importance measure for the feature $f$ is defined as:
\begin{equation}
	M(f)=\sum \frac{(e_{2}-e_{1})}{N_{t}}
\end{equation}

We can assume a condition to explain why this value can represent the importance of the feature. When the random noise is added, the accuracy of the classification of OOB drops significantly. It indicates that this feature has a great influence on the prediction results, thus it is relatively important for classification. By this method, we rank the features according to the value of the importance and retain the features with higher value. Then an effective feature subset is selected.

\subsection{Ensemble Learning}

\begin{figure}[t]
	\center
	\includegraphics[width=0.9\linewidth]{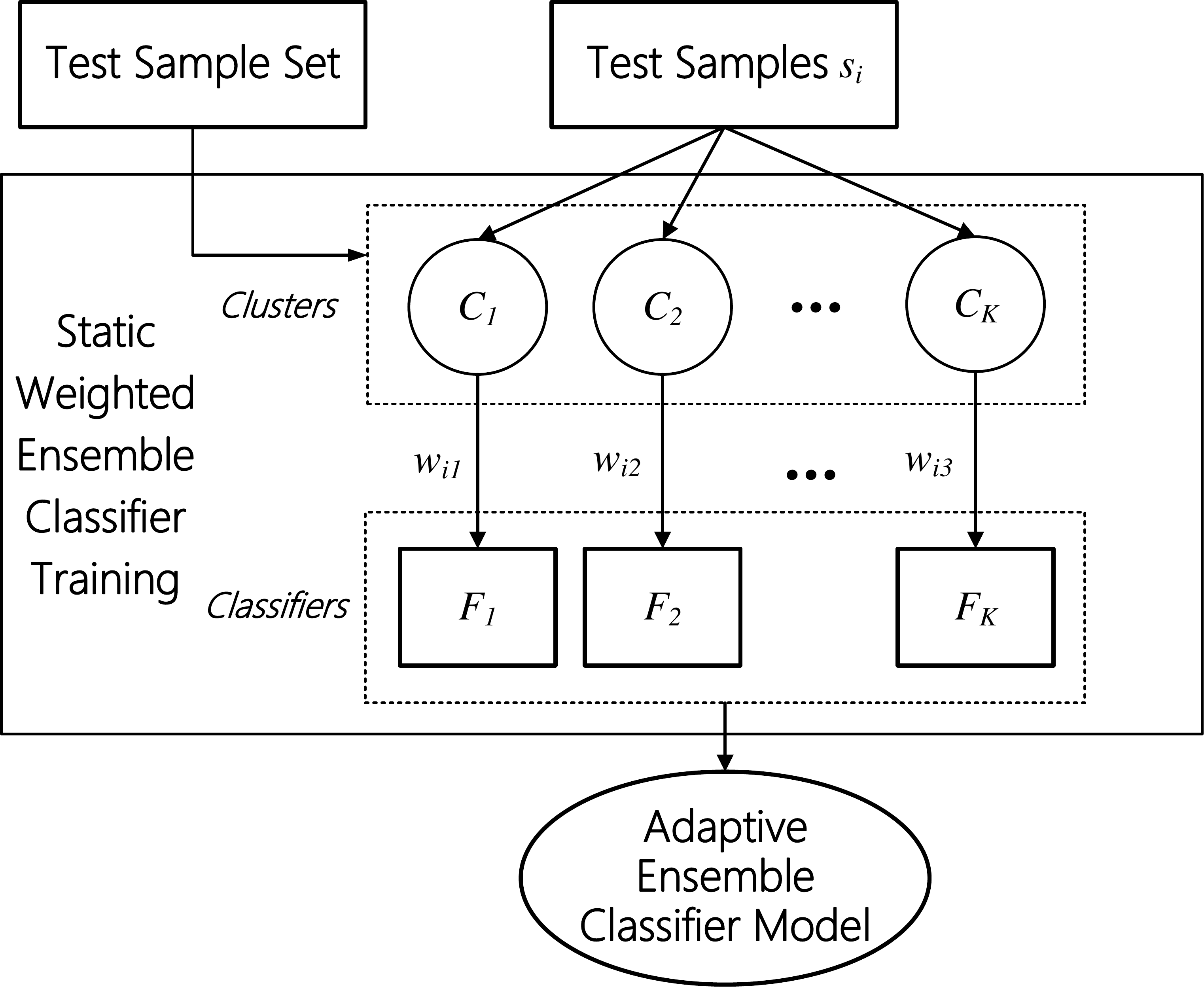}\\
	\caption{The framework of adaptive weighted ensemble classifier}
	\label{Fig.3}
\end{figure}

The purpose of this step is to obtain an ensemble classifier model that can assign the weight between each testing sample and each classifier adaptively for Android malware family classification. To achieve our purpose, we design an adaptive weighted ensemble classifier based on a clustering algorithm and a static weighted ensemble classifier as shown in Fig. \ref{Fig.3}. The clustering algorithm is used to retain the difference of each sample groups and form the base classifiers respectively. Adding weights become an essential part during both of the two phase in ensemble learning. To assure a satisfactory clustering result with an appropriate classifier center and get a high local classification accuracy, we improve the K-means algorithm by adding the weight of each training sample. Additionally, to keep a high classification that do not change with the specificity of a sample, during the testing phase, the weights of base classifiers are dynamically adjusted based on the distances from the testing samples to the sample grouping centers.

\subsubsection{clustering Algorithm}
The execution of the clustering algorithm is unsupervised, which can divide the samples from the data set into several disjoint subsets. The k-means is currently one of the most popular clustering algorithms in the field of Android malware detection and Android malware family classification. Its implementation method is to select the cluster centers and then divide the specified data into the closest clusters from the cluster center through calculating the Euclidean distance formula. In this process, repeated iterations are required until the newly selected cluster center is not significantly different from the previous one. However, the k-means algorithm still has two shortcomings. One is that the k-means algorithm is too dependent on the initialization of the center. Randomly selecting the initial cluster center is easy to make the clustering result fall into the local optimum or be greatly affected by the outlier. Another reason is that when we use Euclidean distance for data partitioning, each feature is considered to have the same effect on classification. According to the phase of feature selection, the importance of each feature for classification is different. We use a method to add weights on the samples by leveraging the importance value of the features calculated by the Random Forest. The Euclidean distance after being weighted can reflect the characteristics of the samples, which lead to a preferable clustering result.

The density-based k-means algorithm is used to obtain a set of clusters and base classifiers. The specific implementation of this algorithm is introduced as follows.

The sample dataset with $n$ features is defined as $S=\{s_1,s_2,...,s_m\}$. Inspired by the importance assigned to each dimension feature by the Random Forest, we normalize these features and assign weights to each dimension feature:
\begin{equation}
	\begin{split}
		&w_{i}=\frac{v_{i}}{\sum_{i=1}^{n}v_{i}}\\
		&\sum_{i=1}^{n}w_{i}=1, 0\le w_{i} \le 1
	\end{split}
\end{equation}
where $v_{i}$ is the value of importance measure assigned to each dimensional feature in the step of the feature selection. So the weighted Euclidean distance formula is defined as:
\begin{equation}
	{d(s_{j},s_{k}})= \sqrt{\sum_{j=1}^{n}w_{i}( s_{i,j}-s_{i,k} )^{2}}
	\label{equ4}
\end{equation}
where $s_{i,j}$ is the $i$th feature in the $j$th sample in the sample set.
We calculate the weighted distance between all samples according to Eqs. (\ref{equ4}). And we calculate the average distance of all the samples as follows: 
\begin{equation}
	Avgdist(S)=\frac{\sum d(s_{p},s_{q}))}{\tbinom{2}{m}}
\end{equation}
where ${\tbinom{2}{m}}$ is the sum of the number of two samples taken from $m$ samples.
Then, we calculate the density parameter of each sample as follows:
\begin{equation}
	Dens(s_{p})=\sum_{q=1}^{m}u(Avgdist(S)-d(s_{p},s_{q}))
\end{equation}
\begin{equation}
	u(z)=
	\begin{cases}
		1 &z > 0\\
		0 &z \le 0
	\end{cases}
\end{equation}
where $Dens(s_{p})$ represents the number of samples within the average distance radius of this sample.

First, the sample with the largest density parameter is selected, and then the sample whose distance is within the average distance of the sample set is removed. By this method, we ensure the diversity of the initial cluster center samples. Repeat the above steps until $k$ initial cluster centers are selected.

According to Eqs. (\ref{equ4}), the remaining samples are divided into the cluster which is closest to the cluster center. Then recalculate the new cluster center as follows:
\begin{equation}
	C(M_{i}) = \frac{1}{\left | M_{i} \right |}\sum_{s_{j}\epsilon M_{i}}s_{j}
\end{equation}
where $C(M_{i})$ is the clustering center, $M_{i}$ is the clustering center and $|M_{i}|$ is the number of samples in the clustering cluster.
We iterate the process of clustering partitioning until the clustering centers are unchanged. 
Finally, all samples can be divided into $k$ clusters.
The improved k-means algorithm based on density is shown in $Algorithm 2$.

\begin{algorithm}[t]
	\small
	\caption{Improved k-means algorithm based on density}
	{\bf INPUT:$~~S:\{s_1,s_2,...,s_m\}$} (set of training samples); k(set of clusters)\\
	1. calculate the distance between all samples:   \\
	$~~~~~{d(s_{j},s_{k}})= \sqrt{\sum_{j=1}^{n}w_{i}( s_{i,j}-s_{i,k} )^{2}}$\\
	2. calculate the average distance: \\
	$~~~~~Avgdist(S)=\frac{\sum d(s_{p},s_{q}))}{\tbinom{2}{m}}$\\
	3. calculate the density of all samples:\\
	$~~~~~Dens(s_{p})=\sum_{q=1}^{m}u(Avgdist(S)-d(s_{p},s_{q}))$\\
	4. \textbf{repeat}\\
	5. ~~~~select the sample with the largest density parameter\\
	$~~~~~~~~$as initial cluster center\\
	6. ~~~~remove the samples within the average distance from\\
	$~~~~~~~~$the sample set\\
	7. \textbf{until} find k initial cluster centers\\
	8. \textbf{repeat}\\
	9. ~~~~the remaining samples are divided into cluster which\\
	$~~~~~~~~$ is closest to the cluster center.\\
	10.~~~~\textbf{for} i=1,2,...,$k$ \textbf{do}\\
	11.~~~~~~~~~~recalculate cluster centers: $C(M_{i}) = \frac{1}{\left | M_{i} \right |}\sum_{s_{j}\epsilon T_{i}}s_{j}$\\
	12.~~~~\textbf{end for}\\
	13. \textbf{until} the clustering centers do not change\\
	{\bf OUTPUT:$C:\{C_1,C_2,...,C_k\}$}(set of clusters)\\
\end{algorithm}

\subsubsection{Adaptive Ensemble Classifier}
We design an adaptive weighted ensemble learning algorithm to predict the testing samples. Static weighted ensemble learning classifiers are mainly divided into two types, boosting and bagging. The boosting-based Adaboost algorithm has been successfully used in Android malware detection with its simple processing process and good detection effect \cite{adroid}. Therefore, we use Adaboost as the static weighted ensemble classifier in our adaptive weighted ensemble learning process.

First, the training samples are clustered according to $Algorithm 2$ to generate $k$ clusters, and the cluster centers are $\{C_{1}, C_{2},..., C_{k}\}$. 
Then the generated $k$ Adaboost classifiers are represented as $\{F_{1}, F_{2},..., F_{k}\}$. 
To get the adaptive weight of each classifier to the testing samples, we calculate the distance from each testing sample to the centroid of each cluster. According to Eqs. (\ref{equ4}), the distance $d_{i,k}$ between the testing sample ${s_{i}}$ and the cluster centroid $C_{k}$ is calculated to generate the distance matrix:\\
\begin{equation}
	Dis = \begin{pmatrix}
		d_{1,1} & \cdots & d_{1,k}\\ 
		\vdots  &  \ddots & \vdots \\ 
		d_{n,1} &\cdots & d_{n,k}
	\end{pmatrix}
\end{equation}
Then, according to Eqs. (\ref{equ10}),\\
\begin{equation}
	w_{i,j}=\frac{(1+d_{i,j})^{-1}}{\sum_{q=1}^{k}(1+d_{i,q})^{-1}}
	\label{equ10}
\end{equation}
the distance matrix is converted into the weight matrix of each cluster classifier for the testing samples:
\begin{equation}
	W = \begin{pmatrix}
		
		w_{1,1} & \cdots & w_{1,k}\\ 
		\vdots  &  \ddots & \vdots \\ 
		w_{n,1} &\cdots & w_{n,k}
	\end{pmatrix}
\end{equation}
The final prediction of the testing sample is obtained according to Eqs. (\ref{equ12}):
\begin{equation}
	\begin{split}
		&P(y_{i} = u) = \sum_{j=1}^{k}P(F_{j}(y_{i}) = u) * w_{i,j})
		\\ 
		&y_{i} = argmax_{u}(P(y_{i} = u)) 
	\end{split}
	\label{equ12}
\end{equation}
In order to calculate the probability of the testing samples being classified into each malware family, we use the weighted summation method to calculate and divide the testing samples into the most likely malware family according to the results.
Through the above steps, the weight of each integrated classifier is supposed adaptive to the testing samples due to the different similarity between each testing sample and each cluster.

\section{Performance and Evaluation}
\subsection{Android Malware Dataset}
In this paper, we use the Drebin dataset \cite{drebin} which is currently the largest public malware mobile dataset with family labels. This malware dataset contains 5560 Android malware from different malicious families, and each malware in the dataset is classified as a certain malicious family. We list the names and the numbers of the ten families with the most samples in Table \ref{table1}.

\begin{table}[t]
	\caption{Android malware dataset distribution}
	\begin{tabular}{ll}
		\hline
		Android Malware Family & Number of Samples \\
		\hline
		FakeInstaller          & 898               \\
		DroidKungFu            & 665               \\
		Plankton               & 623               \\
		Opfake                 & 590               \\
		GinMaster              & 338               \\
		BaseBridge             & 323               \\
		Iconosys               & 150               \\
		Kmin                   & 147               \\
		Fakedoc                & 75               \\
		Geinimi                & 92                \\
		Total                  & 3901             \\
		\hline
	\end{tabular}
	\label{table1}
\end{table}

\subsection{Experiment Setup and Results}
This section shows the experiment setup and presents our experimental results, including the performance of family classification and some significant comparisons.

The computing environment we use to evaluate the performance is a PC installed with Microsoft Windows 10 (64-bit) powered by 3.80-GHz AMD Ryzen Threadripper 3960$\times$ 24-Core Processor and 128-GB RAM. Before our experiments, we use the Apktools (v2.4.0) to decompile the apk files. Then we use the Anaconda (python 3.6) to accomplish the whole process of our experiments. To generate a model of high accuracy of classification, the scikit-learn package \cite{sklearn} is used as our machine learning tool. 

At the start of our experiments, we decompile the apk files by using Apktools to convert them into manifest files and smali files. To obtain a more scientific average result in the experiments, 5-fold cross-validation method is used to divide the dataset into the training set and the testing set. It first divides the experimental samples into 5 sample sets. Each group of samples is selected as the testing set, and the remaining 4 groups of samples will be used as the training set. Then the method use the average result of these 5 experiments as the final result of the experiment. This is a popular method to divide the dataset, \cite{AOM,RSO} adopted the 5-fold cross-validation to evaluate the performance of their approaches on the testing set.

At the end of our experiments, to evaluate the effectiveness of our scheme, we introduce the following evaluation indicators: Accuracy, Macro Precision, Macro Recall, and Macro F1-Score. Accuracy is the proportion of all correctly predicted samples to the total samples. Precision is the proportion of the samples that are correctly predicted to be positive to all samples that are predicted to be positive. Recall is the proportion of the samples that are predicted to be positive to all samples that are truly positive. F1-Score is the harmonic average of precision and recall. Macro precision and macro recall are the average values of precision and recall, and macro F1-Score is the harmonic average value of macro precision and macro recall. The experiments show our proposed scheme achieve the accuracy of 98.92$\%$, the macro precision of 99.23$\%$, the macro recall of 99.17$\%$ and the macro F1-Score of 99.12$\%$.

\subsubsection{Selection of Valid APIs }

\begin{figure}
	\center
	\includegraphics[width=1\linewidth]{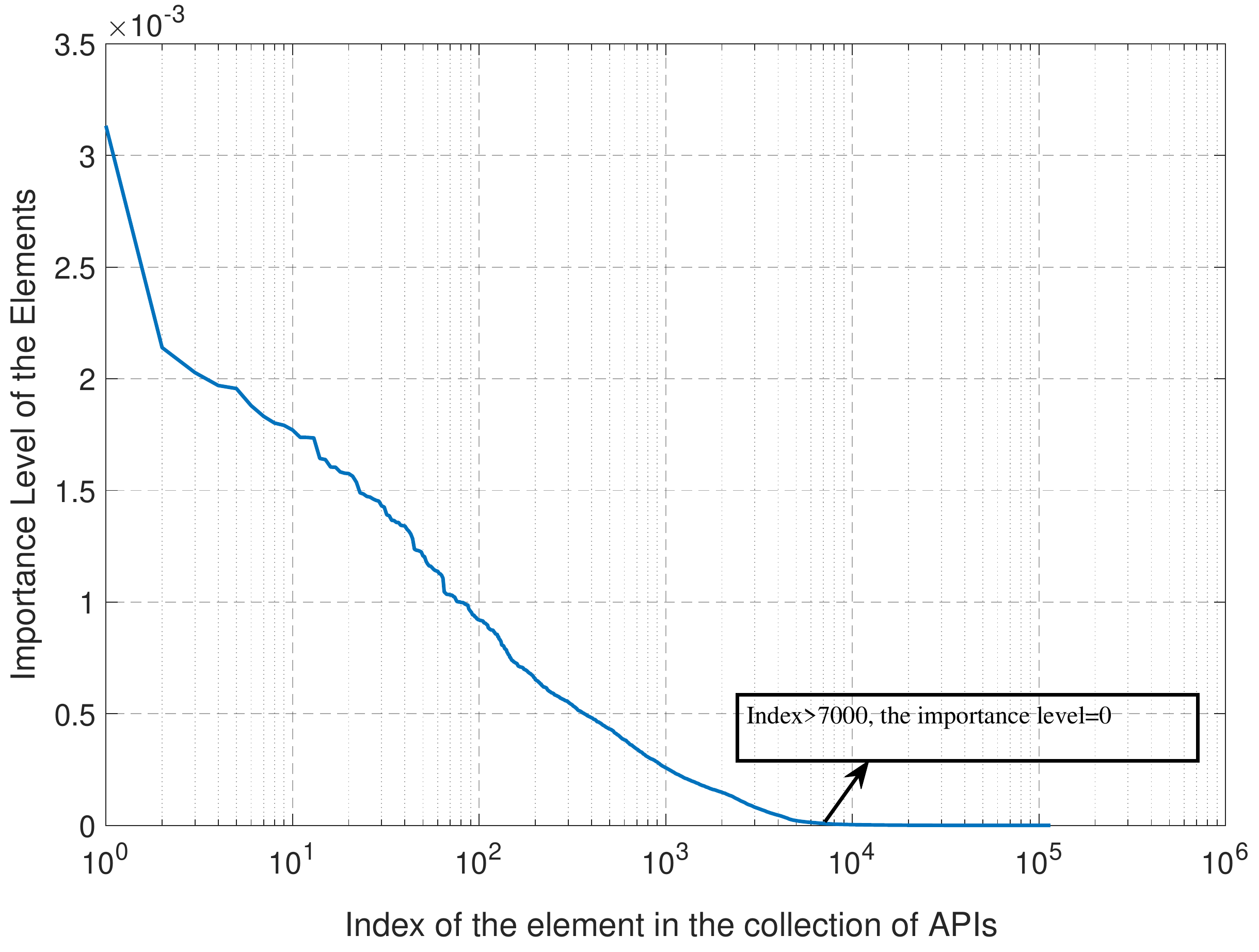}\\
	\caption{Description of the importance of the APIs}
	\label{Fig.4}
\end{figure}

To extract a type of effective implicit features, we leverage the specific combinations of APIs to represent the API call relationship. However, most of the APIs are invalid that badly influence the efficiency of the extraction of the API call relationship. In order to reduce the size of the storage space occupied by the matrix while maintaining the stability and efficiency of our scheme, we need to filter out the useless APIs. We use the Random Forest \cite{random_forest} to evaluate the importance of features. The calculation method of importance value is shown above. As shown in Fig. \ref{Fig.4}, the importance of the APIs ranked after 7000 becomes almost zero. It is found that most of the APIs have almost zero importance value for classification. If these APIs are not filtered out, they will inevitably produce a sparse matrice with quantities of zero elements. Hence, we choose the top 7000 ones as the basic APIs to generate the API call graph and API call relationship matrix. 

\subsubsection{Selection of Proper Features}
\begin{figure}
	\center
	\includegraphics[width=1\linewidth]{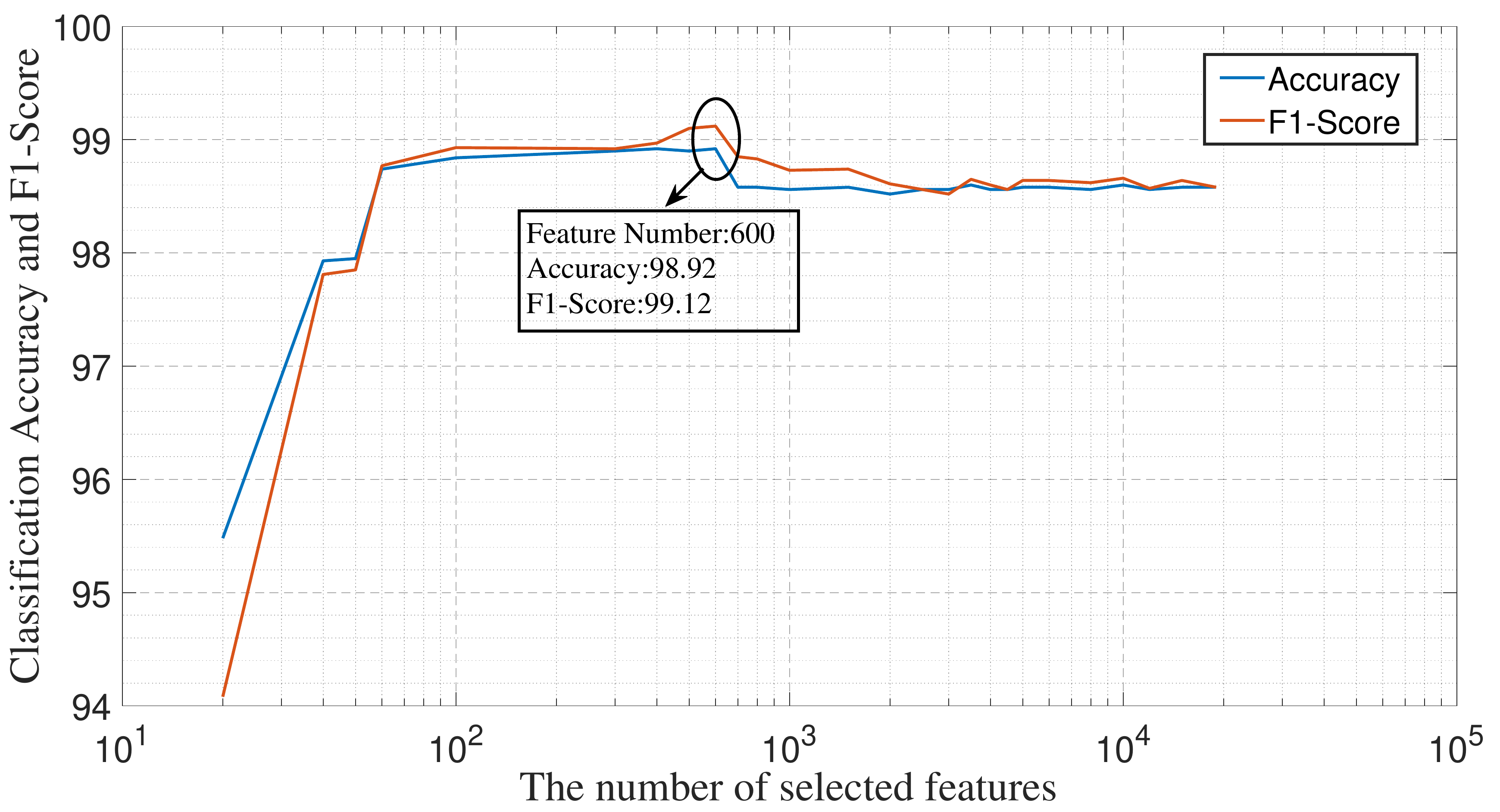}\\
	\caption{Accuracy and F1changes under different API quantities}
	\label{Fig.5}
\end{figure}

Similar to the selection of the valid APIs, the selection of the features is essential to our classification. It not only reduce the data storage space but also achieve better classification results. Before entering the training phase, the selected features of various categories are combined and made into a final feature set list. For further selection, the Random Forest is used to select the features that are most conducive to classification. By this way, the relatively unimportant features are filtered out and we retain the set of features having the best classification efficiency. The classification effects of the experiments are comprehensively evaluated by two indicators including the accuracy and the macro F1-Score. The accuracy rate can directly reflect the correctness of the classification, and the macro F1-Score can reflect both the value of precision and recall. After the experiments , we draw the curve of the accuracy and the F1-Score under different numbers of the selected features. As shown in Fig. \ref{Fig.5}, when the number is 600, both the accuracy and the F1-score reach their maximum values, the accuracy is 98.92$\%$ and the F1-Score is 99.12$\%$. If the number of selected features continues to increase, both the accuracy and the F1-Score will decrease slightly and then tend to stabilize. Therefore, we consider the number 600 as the most suitable number for selected features. After the selection, we obtain the representative and distinctive features that are helpful to the classifier to distinguish malware.

\subsubsection{Malware Family Classification Results}
\begin{figure}
	\center
	\includegraphics[width=1\linewidth]{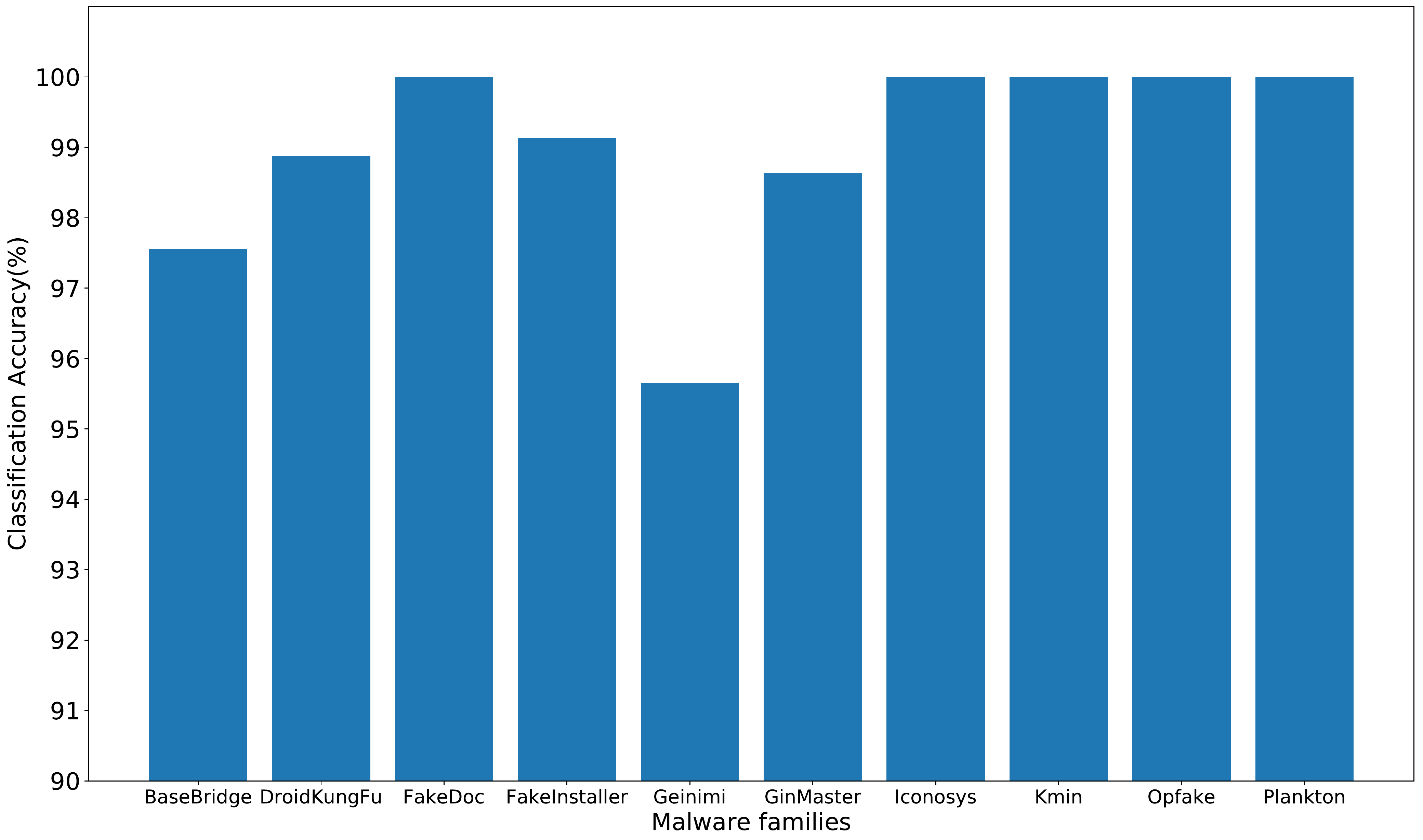}\\
	\caption{Prediction accuracy in each malware family}
	\label{Fig.6}
\end{figure}

\begin{figure}
	\center
	\includegraphics[width=1\linewidth]{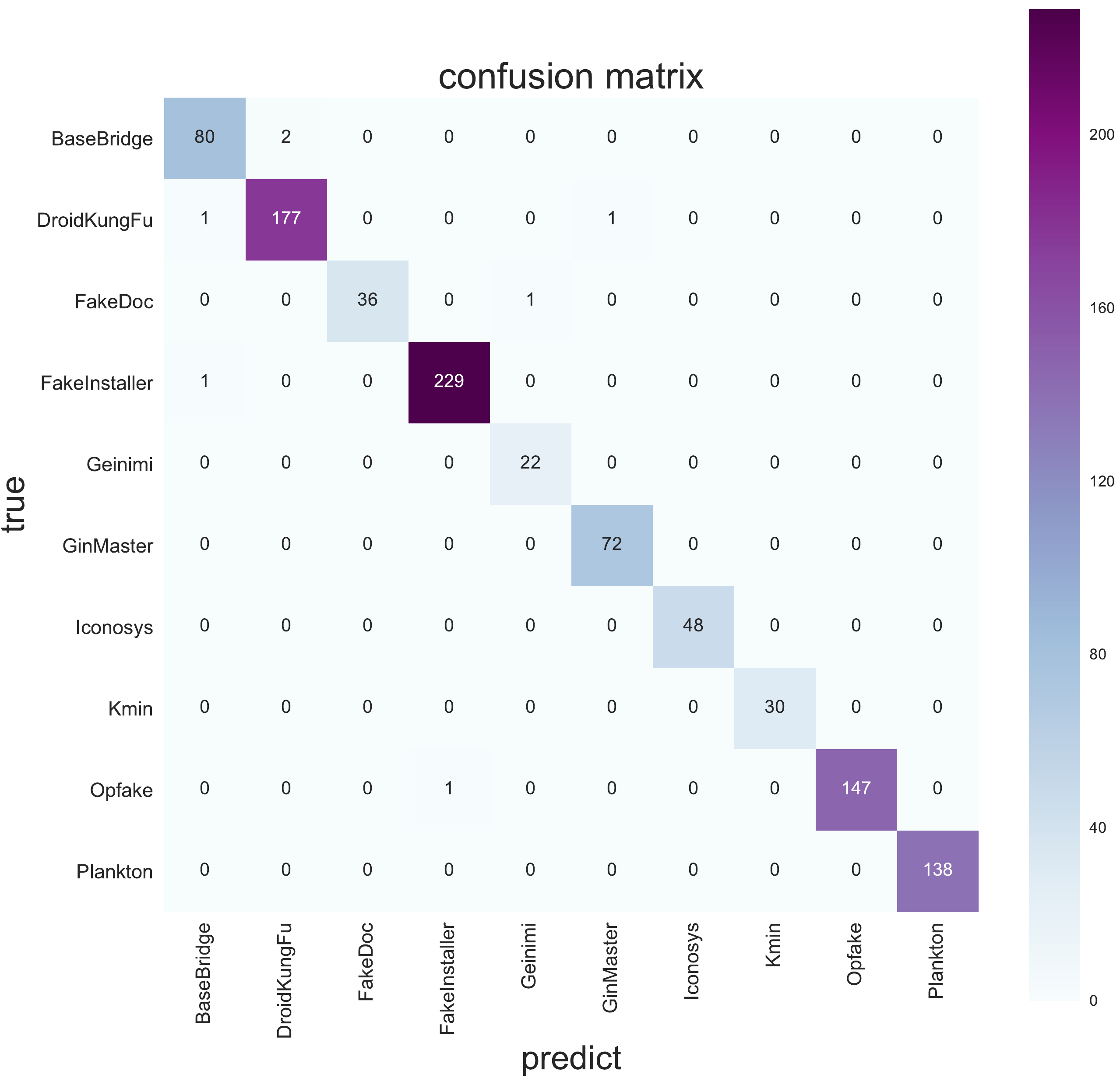}\\
	\caption{Confusion matrix of classification test on Drebin dataset}
	\label{Fig.7}
\end{figure}

For multi-classification of malware families, we use the samples from the Drebin dataset to evaluate the classification results. We list the experiment results of ten largest malware families in the Drebin and compare their accuracies in Fig. \ref{Fig.6}. It can be seen that among these ten families, FakeDoc, Iconosys, Kmin, Opfake, Plankton all achieve the highest 100$\%$ accuracy, even for the family with the lowest accuracy, its accuracy can surpass 95$\%$. Fig. \ref{Fig.7} is the confusion matrix, which is used to give a quick view of the performance of the classification test on each malware family. It shows the number of correct classifications and wrong classifications for each family.

The elements of the matrix reflect the concrete classification result. We can get the similarity and the difference between each malware family by analyzing the matrix. Fig. \ref{Fig.7} proves that our classifier performs a wonderful result. As a whole, most of the samples can be classified as what they belong to. For instance, FakeDoc, Iconosys, Kmin, Opfake and Plankton achieve extremely high accuracy with no error. Nevertheless, there still exists several misclassification. For instance, among 82 BaseBridge's samples, 2 samples which belong to BaseBridge are misclassified as DroidKungFu. The reason is that they have similar behaviors, and therefore similar code, such as permission requests and API calls. They both gain access to the phone to conduct some malicious behaviors that take up resources or endanger the privacy of users by collecting information and then sending it to a remote server. The similarity between their behaviors and codes makes it harder to tell them apart.

\subsubsection{Comparison with 4 Traditional Machine learning Classification Models}

To further demonstrate the superiority of FamDroid, we also use a variety of classical algorithms for comparison. We use 5 traditional machine learning classification models: Support Vector Machine(SVM), Logistic Regression(LR), Gradient Boost Decision Tree(GBDT), Naive Bayes(NB) and K-Nearest Neighbor(K-NN) to evaluate the performance of our selected features and compare them with FamDroid. After sufficient experiments, all 5 models were optimized to their best parameter settings.

\begin{table}[t]
	\caption{Comparison between different algorithms and selected features}
	\begin{tabular}{lllll}
		\hline
		model	& ACC(all features)  & F1(all features)	& ACC(apis)	& F1(apis) \\
		\hline
		AWEC	&0.989	&0.991	&0.968	&0.962\\
		SVM	&0.958	&0.967	&0.922	&0.933\\
		LR	    &0.961	&0.958	&0.931	&0.918\\
		KNN	&0.954	&0.956	&0.937	&0.948\\
		GBDT	&0.972	&0.971	&0.936	&0.924\\
		NB	    &0.951	&0.942	&0.894	&0.868
		\\
		\hline
	\end{tabular}
	\label{tablell}
\end{table}

Table \ref{tablell} shows the accuracy and the F1-score of our classifier and 5 traditional individual machine learning classifier models under the comprehensive combined feature set which consists of a type of implicit features and 4 types of explicit features. While entering the same feature set, the accuracy and the F1-score of the adaptive weighted ensemble classifier are both higher than other machine learning classifier models. FamDroid can reach 0.989 on accuracy and 0.991 on F1-score under the comprehensive combined feature set. In contrast, GBDT with the highest accuracy and F1-score among the classical algorithms reaches 0.972 on accuracy, 0.17 lower than the adaptive weighted ensemble classifier and 0.971 on F1-Score, 0.22 lower than the adaptive weighted ensemble classifier. It can be seen that our classifier has the best performance among all the indicators. Since FamDroid improves the K-means to cluster the training samples and design an algorithm to adaptively assign the weight to each classifier according to the similarity between different samples and clusters, we can keep a high accuracy of the samples from different malware families.  When we use either one of these algorithms under the condition of entering comprehensive combined feature set proposed in this article, the classification effect will be significantly better than the simple feature set which only contains APIs. The reason is that we consider different types of features including the API call relationship that is representative to the common malicious behaviors in a malware family, which makes our classifier easier to classify the samples.

\subsubsection{Comparison with 5 State-of-the-Art Reference Schemes}

\begin{figure}
	\center
	\includegraphics[width=1\linewidth]{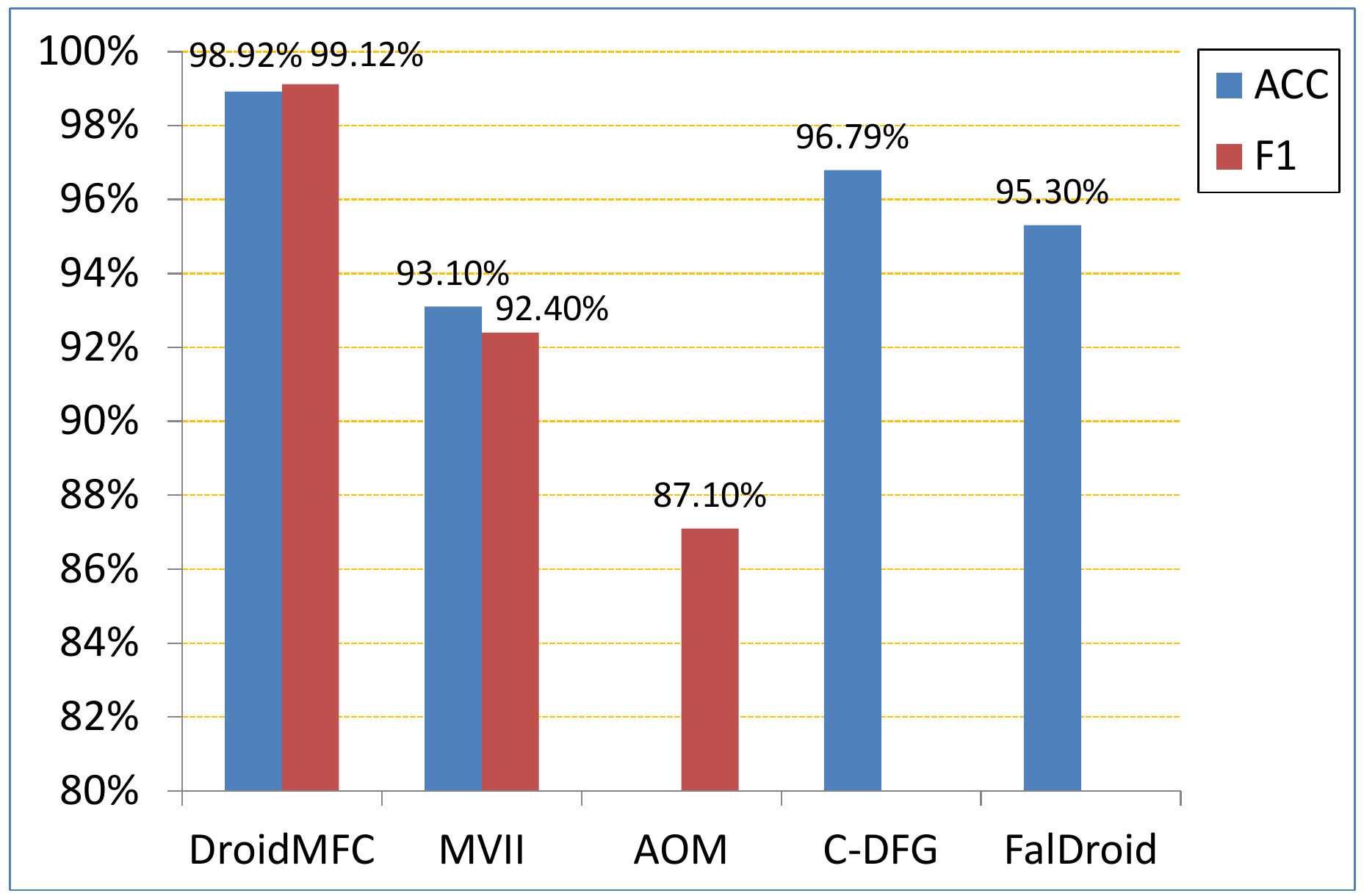}\\
	\caption{Comparsion of state-of-the-art schemes}
	\label{Fig.8}
\end{figure}

To validate our efforts on the improvements of state-of-the-art schemes, we choose 4 reference schemes to compare with FamDroid. MVII (MVIIDroid) \cite{MVIIDroid} uses a Multiple View Information Integration Approach for Android malware detection and family identification. The authors evaluate MVIIDroid using ten-fold cross-validation on the Drebin dataset. MVIIDroid achieves an accuracy of 0.931 and an F1-Score of 0.924. AOM(Android-Oriented Metrics) \cite{AOM} uses Android-Oriented Metrics to identify Android malware families. The authors evaluate AOM on 2,869 malware apps, across 35 different malware families and achieve  a false positive rate of 1.2$\%$ and a F1-Score of 87.1$\%$. C-DFG \cite{C-DFG} uses the horizontal combination of CFG and DFG as features offers. The authors evaluate C-DFG on the Drebin dataset and achieve an accuracy of 96.79$\%$. FalDroid \cite{FalDroid} makes Android malware family classification and representative sample selection via Frequent Subgraph Analysis. They evaluate FalDroid on the Drebin datset with 132 families and achieve 0.953 on classification accuracy. Fig. \ref{Fig.8} shows the comparisons between our scheme and the 4 reference schemes. It can be seen that FamDroid evaluate a better performance than other schemes of the malware family classification on the Drebin dataset.

These schemes mainly contribute to the extraction of efficient features. For malware family classification, what we need do is more than to find sensitive APIs or other features that are helpful to detect malware. Each malware family has its own characteristic and regular when it conducts malicious behaviors. We concentrate on how to extract the type of features to describe the characteristic. Among the above schemes, FalDroid and C-DFG have some similarities with our scheme. FalDroid proposed a weighted-sensitive-API-call-based graph in feature extraction and C-DFG extracted features from control-flow graph and data-flow graph. However, they didn't delve further into the correlation between different API calls and malicious behaviors. FamDroid considers traversing all paths of API calls to generate the API call graph and extracts the call relationship from it. Verified by the experiment, extracting the call relationship between different APIS is a more efficient way in malware family classification. What's more, different from a binary classification, the classifier needs to be effective and with strong adaptability. In contemporary multi-classification research, either a single classifier or a static ensemble classifier is used. The classification effect can be further improved by automatically assigning the weights of the integrated classifier according to the characteristics of the samples, which is a good solution to the multi-classification problem of malicious families. Verified by the experiment, FamDroid uses an adaptive weighted ensemble classifier to obtain a higher accuracy and F1-Score.

\section{Conclusion}
In this paper, we propose a methodology for Android malware family classification. The proposed learning-based approach is based on the static analysis which uses the combination of Permissions, App Components, Intent Filters and API Call Relationship as features and an adaptive weighted ensemble learning algorithm to classify malware families on the Drebin dataset. We carry out a series of experiments and find that our scheme has great advantages in Android malware family classification.

The superiority of FamDroid is mainly embodied in the following two aspects. (1) We not only use the traditional classification-friendly feature set but use an implicit feature set including the API call relationship which perfectly represents the common characteristics of malware families. In the implicit feature extracting method, we select the APIs with high importance value and use them to generate an feature matrix which stores the API call relationship. The features selected from the matrix embodies the most risky and characteristic behaviors to a malware family. (2) We propose an adaptive weighted ensemble learning classifier which contains the clustering algorithm based on the improved k-means and the improved ensemble classifier by adding the weight between each testing sample and each base classifier. The clustering for the samples and adaptive weight allocation make a satisfactory multi-classification result. Since our research works on the two aspects , FamDroid can achieve a high accuracy of 98.92$\%$ and a high F1 score of 99.12$\%$, better than other reference schemes.

\section*{Acknowledgment} \small
This work is supported in part by National Natural Science Foundation of China (Grant No. 61821001), and Fundamental Research Funds for the Central Universities.

\ifCLASSOPTIONcaptionsoff
  \newpage
\fi



%
\bibliographystyle{ieeetr}
\bibliography{references}

\end{document}